\documentclass{article}

\author{Edward Kapu\'scik\footnote{email:Edward.Kapuscik@ifj.edu.pl}\\ The Alfred Meissner Graduate School \\ for Dental Engineering and Humanities\\ Ustro\'n, ul. S³oneczna 2, Poland}

\title{Special Theory of Relativity without special assumptions and tachyonic motion}

\bigskip
\date{\textit{Dedicated to N. N. Bogolyubov, Jr.}\\ 
\bigskip
\noindent PACS 0.3.30.+p}
\begin{document}
\maketitle

\begin{abstract}
The most general form of transformations of spacetime coordinates in Special Theory of Relativity based solely on physical assumptions are described. Only the linearity of spacetime transformations and the constancy of the speed of light are used as assumptions. The application to tachyonic motion is indicated.
\end{abstract}

\section*{Introduction}

In almost all textbooks of Special Relativity \cite{1} it is claimed that  the fundamental
transformations of spacetime coordinates $\left(\vec{x},t\right)$ have the form of the Lorentz transformations

$$t\rightarrow t'=\gamma\left(t-\frac{\vec{V}\cdot\vec{x}}{c^2}\right),\eqno(0.1)$$
$$\vec{x}\rightarrow\vec{x}'=\vec{x}+(\gamma-1)\frac{(\vec{V}\cdot\vec{x})}{V^2}\vec{V}-\gamma\vec{V} t,
\eqno(0.2)$$
where 
$$\gamma=\left(1-\frac{V^2}{c^2}\right)^{-1/2}
\eqno(0.3)$$
is the famous Lorentz factor with $\vec{V}$  being the relative velocity of two observers and $c$ is the velocity of light.

It turns out however that \textbf{such statement is not quite true}. In
fact, historically \cite{2} H. A. Lorentz derived his
transformations from the requiring of the
covariance of the vacuum Maxwell equations under
linear transformations of spacetime
coordinates. 
Now it is  known that in any medium Maxwell 
equations are covariant not only under linear
transformations but 
they are covariant under arbitrary
transformations of spacetime coordinates, as well. Therefore, the Lorentz derivation of Lorentz transformation used additional particular assumption that the basic Maxwell equations are the vacuum Maxwell equations.  

Also Einstein\cite {3}, in his derivation of the
spacetime transformations, used particular
information on the Doppler effect and assumed
the equality of  two-ways velocities of light. Again, we see the additional particular assumptions.
It is not difficult to show that without these assumptions we can get more general forms of Lorentz transformations \cite{4}.

In modern textbooks \cite {5} the Lorentz transformations are
usually derived from the assumption that
they leave invariant the Minkowski interval
$$ds^2=c^2(dt)^2-(d\vec{x})^2.
\eqno(0.4)$$

\textbf{It is not difficult to understand that all such approaches
use additional assumptions without sufficient
physical justification !}

In the present paper, which I dedicate to Professor N. N. Bogolyubov, Jr., we shall present a new derivation of fundamental transformations of spacetime coordinates based solely on physical assumptions. As a result we get a slightly more general form of fundamental transformations of spacetime coordinates. On the basis of that new form of Lorentz transformations we show that in the framework of Special Relativity we may equally well describe both sub- and superluminal motions.

\section{Physical Approach to Special Relativity}

To begin with, let us remind that Special Relativity is
based only on two physically justified
assumptions:

\begin{itemize}
\item The uniform motion is an invariant notion.

\item There exists an invariant velocity. 
\end{itemize}

From the first assumption we have the
transformations:
$$t\rightarrow t'=At+B_k x^k,\eqno(1.1)$$
$$x^k\rightarrow x'^k=D^k_j x^j +E^k t\eqno(1.2)$$
with some yet unknown coefficients $A, B_k, D_j^k$ and $E^k$. Here the summation over repeated indexes is understood. 

From these transformations we get the
following transformation rule for velocities of
motions
$$v^k(t)=\frac{dx^k(t)}{dt}\rightarrow v'^k(t')=\frac{dx'^k(t')}{dt'}=\frac{D^k_j v^j(t)+E^k}{A+B_k v^k(t)}.\eqno(1.3)$$
The requirement of the existence of an invariant 
velocity $\vec{C}$ means that the magnitude of this velocity must be the same in each reference frame. Therefore for the velocity $\vec{C}$ we have
$$\vec{C}'^2=\vec{C}^2\eqno(1.4)$$
and in view of the transformation rule (1.3) for velocities this leads to the relation 
$$\vec{C}^2\left[A+B_kC^k\right]^2=\sum_{k=1}^3\left[D^k_j C^j+E^k\right]\left[D^k_l C^l+E^k\right].\eqno(1.5)$$
We may treat this relation as an equation for the square of the invariant velocity $\vec{C}^2$ which \cite {4}
may have no
solution, one solution or two solutions which
may be of the same modulus or quite
different. Let us consider the case when the
solution of (1.5) is the square of the usual 
velocity of light. In the more general case eq. (1.5)
may have two solutions which correspond to the inequality of two-way velocities of light. Eq.(1.5) leads to some conditions on the parameters  $A, B_k, D_j^k$ and $E^k$. Solving these conditions as a result we arrive to the following general form of the spacetime transformations
$$
t\rightarrow t'=At+\vec{B}\cdot\vec{x},\eqno(1.6)$$
$$\vec{x}\rightarrow\vec{x}'=\sqrt{A^2-c^2\vec{B}^2}(\vec{\textit{R}\vec{x}})+\frac{A-\sqrt{A^2-c^2\vec{B}^2}}{\vec{B}^2}(\vec{\textit{R}B})(\vec{B}\cdot{x})+c^2(\vec{\textit{R}B})t.\eqno(1.7)$$
Here $A$ and $\vec{B}$ are arbitrary parameters and $\textit{R}$ denotes an arbitrary orthogonal matrix.
The transformations make sense only if
$$
A^2-c^2\vec{B}^2>0.\eqno(1.8)$$
As we shall see below, this condition leads to the restriction on velocities of motion.

The standard Lorentz transformations (0.1) and (0.2) are obtained with the choice
$$A=\gamma, \ \ \ \vec{B}=-\frac{\gamma}{c^2}\vec{V}, \ \ \ R=I.\eqno(1.10)$$

In the two-dimensional spacetime the transformations (1.6) and (1.7) reduce to
$$t'=At+Bx,\eqno(1.11)$$
$$x'=Ax+c^2Bt.\eqno(1.12)$$
For shortness we shall refer to our form of Lorentz transformations as the $A,B,C$ form of Lorentz transformations.

In $N$-dimensional spacetime the Jacobian of the generalized Lorentz transformations is equal to
$$
J=\left(A^2-c^2\vec{B}^2\right)^{N/2}.\eqno(1.13)$$
Correspondingly, in the two dimensional spacetime we have
$$
J=A^2-c^2B^2,\eqno(1.14)$$
in three-dimensional spacetime 
$$
J=\left(A^2-c^2\vec{B}^2\right)^{3/2}\eqno(1.15)
$$
and
$$
J=\left(A^2-c^2\vec{B}^2\right)^2\eqno(1.16)
$$
in the four-dimensional spacetime. 

The $A,B,C$ transformations form a group with the following group composition law
$$A_{2,1}=A_2 A_1+c^2(\vec{B}_2\cdot\vec{B}_1),\eqno(1.17)$$
$$\vec{B}_{2,1}=\sqrt{A_1^2-\vec{B}_1^2c^2}R^{-1}_1\vec{B}_2+\frac{A_1-\sqrt{A_1^2-\vec{B}_1^2c^2}}{\vec{B}_1^2}(\vec{B}_2\cdot R_1\vec{B}_1)\vec{B}_1+	A_2\vec{B}_1.\eqno(1.18)
$$
The composition law for space rotations is much more complicated. In fact, repeating twice the spacetime transformations (1.6) and (1.7) we arrive to the composed rotation matrix $R_{21}$ of the form
$$
R_{21}=\Omega_{21}(A_2, \vec{B}_2, R_2 ; A_1, \vec{B}_1, R_1)R_2R_1,
\eqno(1.19)
$$
where $\Omega_{21}$ is an orthogonal matrix which satisfies the following equation
$$
\sqrt{A_2^2-c^2\vec{K}}\vec{K}+\frac{A_2-\sqrt{A_2^2-c^2\vec{L}}}{\vec{L}^2}(\vec{K}\cdot{L})\vec{L}+A_1\vec{L}=
$$
$$ =\Omega_{21}\left[\sqrt{A_1^2-c^2\vec{K}^2}\vec{L}+\frac{A_1-\sqrt{A_1^2-c^2\vec{K}^2}}{\vec{K}^2}(\vec{K}\cdot\vec{L})\vec{K}+A_2\vec{K}\right].
\eqno(1.20)
$$  
Here
$$
\vec{K}=R_2R_1\vec{B}_1, \ \ \ \ \ \vec{L}=R_2\vec{B}_2.
\eqno(1.21)
$$
Unfortunately, it is not easy to solve equation (1.20) and find an explicit form of $\Omega_{21}$. It certainly is an orthogonal matrix because the lengths of vectors on both sides of (1.20) are the same and equal to
$$
(A_2\vec{K}+A_1\vec{L})^2+c^2[(\vec{K}\cdot{L})^2-\vec{K}^2\vec{L}^2].
\eqno(1.22)
$$
From (1.20) it is clear that the vectors in both sides of (1.20) lie in the plane spanned by the vectors $\vec{K}$ and $\vec{L}$. The rotation $\Omega_{21}$ therefore is around the axis parallel to $\vec{K}\times\vec{L}$. The angle of this rotation may be calculated from the scalar product of the vectors in both sides of (1.20).

In standard treatments of Special Relativity the problem of combining two rotations is not well-understood \cite{7} and there is a widely accepted erroneous suggestion that a correction of the composition law of two Lorentz transformations by an additional rotation is needed. As a matter of fact our treatment of Lorentz transformations shows that no correction of the group theoretical composition law is needed. The correct statement is that Lorentz transformations without rotations do not form a subgroup of the group of spacetime transformations because the composition of two Lorentz transformations with $R_1=R_2=I$ is a Lorentz transformation with $R_{21}\ne I$. It is the specific property of Lorentz transformation that always an additional relativistic rotation appears.

The neutral element for the composition laws (1.17) and (1.18) is given by $A=1, \ \vec{B}=0$, \ R=I. The parameters of the reverse transformation are given by
$$A^-=\frac{A}{A^2-\vec{B}^2c^2},\ \ \  \vec{B}^-=-\frac{\vec{RB}}{A^2-\vec{B}^2c^2} \ \ \  R^-=R^{-1}.\eqno(1.19)
$$

\section{Velocities of motion}

From the transformation rules (1.6) and (1.7) for spacetime coordinates we get the following transformation rule for velocities of motion
$$\vec{V}'=\frac{\sqrt{A^2-c^2\vec{B}^2}(R\vec{V})+\frac{A-\sqrt{A^2-c^2\vec{B}^2}}{\vec{B}^2}(R\vec{B})(\vec{B}\cdot\vec{V})+c^2(R\vec{B})}{A+\vec{B}\cdot\vec{V}}.\eqno(2.1)$$

Assuming that there exist a reference frame (for which we choose the unprimed coordinates) in which a physical object is at rest we find that in the transformed reference system this object moves uniformly with the velocity (we omite the unnecessary prime on the left hand side)
$$
\vec{V}=\frac{c^2}{A}R\vec{B}.\eqno(2.2)
$$
Using this relation in the formula for Lorentz transformations (1.6) and (1.7), we may replace $\vec{B}$ by $\frac{A}{c^2}R^{-1}\vec{V}$ and arrive to the following form of the Lorentz transformations

$$
t'=A\left[t+\frac{(\vec{V}\cdot R\vec{x})}{c^2}\right],\eqno(2.3)
$$
$$
\vec{x}'=A\left[\sqrt{1-\frac{\vec{V}^2}{c^2}}(R\vec{x})+\left(1-\sqrt{1-\frac{\vec{V}^2}{c^2}}\right)\frac{\left(\vec{V}\cdot R\vec{x}\right)}{\vec{V}^2}\vec{V}+\vec{V}t\right].\eqno(2.4)
$$
This is almost the same as the standard Lorentz transformation with except of the arbitrary factor $A$. Assuming that this factor is a function of the velocity $\vec{V}$ the composition rule  
$$
A(\vec{V}_{21})=A(\vec{V}_2)A(\vec{V}_1)\left(1+\frac{\vec{V}_1\cdot\vec{V}_2}{c^2}\right)\eqno(2.5)
$$
becomes a functional equation for the function $A(\vec{V})$ with the only solution
$$
A=\gamma.\eqno(2.6)
$$
Here, of course, $\vec{V}_{21}$ denotes the relativistic composition of the velocities $\vec{V}_1$ and $\vec {V}_2$. We stress however the fact that the choice (2.6) is not dictated by Special Theory of Relativity.

It is clear that, using (2.2), from the positivity condition $A^2-c^2\vec{B}^2>0$ we get that
$$
\vec{V}^2<c^2.\eqno(2.7)
$$  

Let us now, contrary to the previous case, assume that in the unprimed reference frame some object moves with an infinite velocity. Then, in the primed reference frame its velocity $\vec{W}$ is equal to
$$
\vec{W}=\frac{A}{\vec{B}^2} R\vec{B}.\eqno(2.8)
$$
Writing the vector parameter $\vec{B}$ in the spacetime transformation (1.6) and (1.7) in terms of this velocity we get

$$
t'=A\left(t+\frac{\vec{W}\cdot R\vec{x}}{\vec{W}^2}\right),\eqno(2.9)
$$
$$
\vec{x}'=A\left[\sqrt{1-\frac{c^2}{\vec{W}^2}}\left(R\vec{x}\right)+\left(1-\sqrt{1-\frac{c^2}{\vec{W}^2}}\right)\frac{\left(\vec{W}\cdot R\vec{x}\right)}{\vec{W}^2}\vec{W}+\frac{c^2}{\vec{W}^2}\vec{W}t\right]\eqno(2.10)
$$
with an arbitrary factor $A$.
Again, assuming that this factor is a function of the velocity $\vec{W}$ from the functional equation which follows from the composition law (1.17) we get
$$
A(\vec{W})=\sqrt{1-\frac{c^2}{\vec{W}^2}}\eqno(2.11)
$$
and from the positivity condition (1.8) we get
$$
\vec{W}^2>c^2.\eqno(2.12)
$$

Therefore, in our version of Relativity Theory we equally well may describe both subluminal and superluminal motions. We must however remember that for the superluminal objects (tachyons) there does not exist the rest reference frame.

It must be also stressed that the velocity of the relative motion of two reference frames is equal to $$c^2\frac{\vec{B}}{A}\eqno(2.13)$$ 
and therefore from the positivity condition (1.8) it is always less than the velocity of light. 

\section{Velocity dependent tensors}

Physical quantities have definite tensorial properties. In Special Relativity a particular role play tensors or pseudotensors which are form-invariant functions of the velocity $\vec{v}$ of moving objects. Pseudotensors of type $(K,L)$ and of rank D on the right hand side of the transformation rule are homogeneous functions of $A$ and $\vec{B}$ of degree $K-L+ND$, where $N$ is the dimension of the spacetime and $D$ the degree of the Jacobian on the right hand side of the transformation rule. The velocity transformation rule (2.1) on the right hand side is a homogeneous function of $A$ and $\vec{B}$ of degree $0$. Therefore, all velocity dependent (pseudo) tensors must satisfy the condition
$$
K-L+ND=0.
\eqno(3.1)
$$
For pure tensors we have $D=0$ and consequently $K=L$. Therefore, only mixed tensors with the same degree of contra- and covariance may be functions of the velocity. The example of such a tensor is  the velocity tensor $V^\mu_\nu(\vec{v})$ discussed in \cite {6}.

It is interesting to note that transformation rules of tensors or pseudotensors which are form-invariant functions of velocities convert into sets of functional equations from which the components of these tensors can be found.  

\section{Four momentum of objects}

  As an application of the presented formalism we shall construct the four-momen\-tum for moving objects. For the contravariant components of the four-momentum we have $K=1,\ \ L=0$. Since momentum is certainly velocity dependent from (3.1) it follows then that in the four-dimensional spacetime we must have $D=-1/4$.
The contravariant components $P^\mu$ of the four-momentum transform therefore as follows
$$
{P'}^0=J^{-1/4}\left[AP^0+\vec{B}\cdot\vec{P}\right],\eqno(4.1)$$
$$\vec{P}'=J^{-1/4}\left[c^2(R\vec{B})P^0+\sqrt{A^2-c^2\vec{B}^2}(\vec{RP})+\frac{A-\sqrt{A^2-c^2\vec{B}^2}}{\vec{B}^2}(\vec{B}\cdot\vec{P})(\vec{RB})\right],\eqno(4.2)$$ 
where $J$ is the Jacobian (1.16).
Assuming that momentum is a form-invariant function of the velocity we get the following set of functional equations for the component of momentum
$$
{P}^0(\vec{V}')=J^{-1/4}\left[AP^0(\vec{V})+\vec{B}\cdot\vec{P}(\vec{V})\right],\eqno(4.3)$$
$$\vec{P}(\vec{V}')=J^{-1/4}\left[c^2(R\vec{B})P^0(\vec{V})+\sqrt{A^2-c^2\vec{B}^2}(\vec{RP}(\vec{V}))+\right.$$

$$\left.+\frac{A-\sqrt{A^2-c^2\vec{B}^2}}{\vec{B}^2}(\vec{B}\cdot\vec{P}(\vec{V}))(\vec{RB})\right],\eqno(4.4)$$
where $\vec{V}'$ is connected with $\vec{V}$ by the transformation rule (2.1).

For the subluminal motions putting in the unprimed reference frame $\vec{V}=0$ the velocity in the primed reference frame is given by (2.2). For the $0$-th component of the momentum we have therefore
$$
P^0(\vec{V})=\frac{1}{\sqrt{1-\frac{\vec{V}^2}{c^2}}}\left[P^0(0)+\frac{\vec{V}\cdot R\vec{P}(0)}{c^2}\right]\eqno(4.5)
$$
and a more complicated expression for the space components of the momentum. Here $P^0(0)$ and $\vec{P}(0)$ are arbitrary constants. Observing that energy should be an even function of velocity we should take $\vec{P}(0)=0$ and then we end up with the four-momentum equal to
$$
P^0(\vec{V})=\frac{P^0(0)}{\sqrt{1-\frac{\vec{V}^2}{c^2}}},\eqno(4.6)
$$
$$
\vec{P}(\vec{V})=\frac{P^0(0)}{\sqrt{1-\frac{\vec{V}^2}{c^2}}}\vec{V}.\eqno(4.7)
$$
To have the standard expression for the space components of momentum we put $P^0(0)=M$, 
where $M$ is the mass of the particle.

In the same way for the covariant components of the four-momentum (for which $K=0,\ \ L=1,\ \ D=+1/4$) we get
$$
P_0(\vec{V})=\frac{P_0(0)}{\sqrt{1-\frac{\vec{V}^2}{c^2}}},\eqno(4.8)
$$
and
$$
\vec{P}(\vec{V})=-\frac{P_0(0)}{c^2}\frac{1}{\sqrt{1-\frac{\vec{V}^2}{c^2}}}\vec{V}.\eqno(4.9)
$$
The standard expression for the momentum we get provided 
$$
P_0(0)=-Mc^2\eqno(4.10)
$$.

For the superluminal motion we have to use the basic transformations of spacetime coordinates in  the form  of (2.9) and (2.10). Proceeding similarly as above we end up with the following expression for  momentum 
$$
P_0(\vec{W})=\frac{P_0(0)}{\sqrt{1-\frac{c^2}{\vec{W}^2}}}
\eqno(4.11)
$$
and
$$
\vec{P}(\vec{W})=-\frac{P_0(0)\vec{W}}{\vec{W}^2\sqrt{1-\frac{c^2}{\vec{W}^2}}}
\eqno(4.12)
$$
However, in this case it is not worthy to write $P_0(0)$ in terms of any mass parameter $M$ as we did in (4.10) because for a tachyon there does not exist the rest frame in which we could measure its mass. From (4.11) and (4.12) we see that for $\vec{W}\rightarrow\infty$ the component $P_0$ remains finite while $\vec{P}$ tends to zero. It is therefore suggestive to assume that $P_0(0)$ is equal to some unknown kind of energy, for example, the dark energy present in the Universe.

\section*{Conclusion}

The present paper creates a firm theoretical basis for tachyonic physics. We have shown that superluminal motions are allowed by Special Relativity in the same way as the subluminal motions are. There is no doubt that Special Relativity does not forbids the existence of tachyons for which there do not exist rest frames.

From the physical point of view the most important problem of experimental discovery of tachyons arises. In the forthcoming paper we shall argue that in the famous experiment by G. Nimtz\cite {8} the tachyons were indeed created in the first part of the paraffin block as a result of the interaction of the incoming photons with phonons. After traveling through the gap between paraffin blocks the tachyons are subsequently annihilated by phonons in the second block with the creation of the final photons.

\end{document}